\documentclass{aastex}          
\usepackage{spr-astr-addons}    

\begin{document}
%
\title{Semiconvective Mixing in Low-Mass Stars}

\shorttitle{Semiconvection in Low-Mass Stars}
\shortauthors{Silva Aguirre et al.}

\author{Silva Aguirre, V.}
\affil{Max Planck Institute for Astrophysics, Garching, Germany} 
\email{vsilva@mpa-garching.mpg.de} 
\and 
\author{Ballot, J.}
\affil{LATT/Observatoire Midi-Pyr\'en\'ees, Toulouse, France}
\and 
\author{Serenelli, A.}
\affil{Max Planck Institute for Astrophysics, Garching, Germany} 
\and 
\author{Weiss, A.}
\affil{Max Planck Institute for Astrophysics, Garching, Germany} 

\begin{abstract}
Mixing processes such as convection, overshooting and rotational mixing have long been known to affect the evolutionary properties of low-mass stars. While modeling a 1.2 M$_\odot$ star, we encountered a semiconvective region outside the fully convective core, reminiscent of the well-known situation for massive stars. In this study we focus on low-mass stars presenting convective cores and, by applying different prescriptions for the determination of the convective boundaries and using different mixing descriptions for the dynamical processes in the core, we look for the effects of semiconvective mixing in the interior structure of the stars and its observable quantities. With this purpose, we have constructed different sets of evolutionary models using a stellar evolution code (GARSTEC), and analyzed the models looking for imprints of these processes. 
\end{abstract}


\section{Introduction}\label{s:1}
It has long been known that different mixing processes in stellar cores strongly influence the overall evolution of stars. The treatment of convectively unstable zones and overshooting regions in evolutionary models affects the outcoming luminosity, temperature, and main-sequence lifetime of the stars, among other consequences.\\
One of the mixing processes that has been extensively studied in massive stars evolution is that of semiconvection. During the hydrogen-burning phase the radiation pressure makes the core expand. Opacities are increased outside the core by electron scattering, where a chemical discontinuity appears as a result of this core expansion. Since the pioneer work of \citet{sh58}, several authors have investigated the occurrence of semiconvective mixing in massive stars and its effects on stellar evolution \citep[e.g.][]{rs70,sc75,nl85}.\\
Although semiconvection was initially thought to occur only in massive stars, it was also found in low-mass stars as a consequence of a discontinuity in the molecular weight produced either by convective core expansion due to the increasing importance of the \textit{CNO} cycle over the \textit{pp} chain \citep{rm72,hs75}, or by the retreating convective core leaving behind a chemical discontinuity \citep{fc73}, both cases producing higher opacities outside the convective core.\\ 
We address the issue of the determination of convective boundaries and the treatment of mixing zones which present a gradient in the molecular weight by the inclusion of the Ledoux criterion for convective instability, and a diffusive approach for semiconvective mixing in our stellar evolution code.

\section{Mixing Prescriptions}\label{s:2}
The convective zone boundaries are defined either with the Schwarzschild criterion \citep{sh58} or the Ledoux criterion \citep{pl47}. For the latter case, the Ledoux temperature gradient is defined as
\begin{center}
\begin{equation}\label{eqn:led}
\nabla_{\mathrm{L}}=\nabla_{\mathrm{ad}}+\frac{\beta}{4-3\beta}\nabla_{\mathrm{\mu}},
\end{equation}
\end{center}
\noindent{where $\nabla_{\mathrm{\mu}}=d\ \mathrm{ln\ \mathrm{\mu}}/d\ \mathrm{ln\ P}$, and $\beta$ is the ratio of gas pressure to total pressure.} If the Ledoux criterion is applied, a zone is considered convectively unstable if $\nabla_{\mathrm{L}}<\nabla_{\mathrm{rad}}$, while in the case of Schwarzschild criterion this reduces to $\nabla_{\mathrm{ad}}<\nabla_{\mathrm{rad}}$. Convective zones can be mixed instantaneously or as a diffusive process using the convective velocity estimated from the mixing-length theory.\\
The inclusion of molecular weight gradients in the criterion for convection allows us to identify semiconvective zones, which are defined as the regions where $\nabla_{\mathrm{ad}} < \nabla_{\mathrm{rad}} < \nabla_{\mathrm{L}}$. Once a semiconvective region has been found, the temperature gradient is calculated using the prescription given by \citet{nl83,nl85}, and a time-dependent diffusive mixing in the semiconvective zone is performed. Both the diffusion coefficient and the temperature gradient in this prescription depend on an efficiency parameter of semiconvection ($\alpha_{\mathrm{sc}}$). We stress the fact that unlike this study, previous efforts to treat semiconvection in low-mass stars \citep[e.g.][]{cm82} considered the mixing in the semiconvective layer to be performed by adjusting the composition until convective neutrality according to the Schwarzschild criterion is reached, making the hydrogen discontinuity in the boundary of the convective core disappear.\\ 
Convective overshooting is also considered in our study and it is implemented as a diffusive process consisting in an exponential decay of the convective velocities within the radiative zone. The diffusion constant is given by
\begin{center}
\begin{equation}\label{eqn:ove}
 D\left(z\right) = D_0 \ \exp \ \left(\frac{-2z}{\xi H_p}\right),
\end{equation}
\end{center}
\noindent{where $\xi$ corresponds to an efficiency parameter calibrated with open clusters, $H_p$ is the pressure scale height, $z$ is the distance from the convective border, and the constant $D_0$ is derived from MLT-convective velocities.} The extent of the overshooting region is limited in the case of small convective cores.

\section{Models}\label{s:3}
\begin{center}
\begin{figure}[t]
\includegraphics[width=\columnwidth]{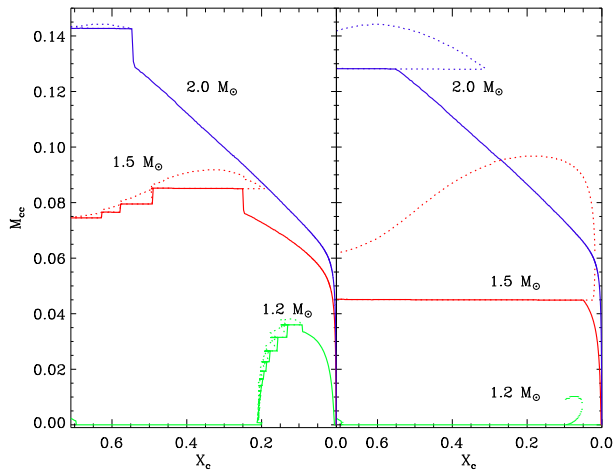}
\caption{\textit{Left panel:} Models computed using the Schwarzschild criterion for convective boundary definition. \textit{Right panel:} Models computed using the Ledoux criterion for convective boundary definition. Solid lines depict the convective core evolution, while dotted lines show the extent of semiconvective zones. No semiconvective mixing or overshooting was applied. See text for details} 
\label{fig:cores}
\end{figure}
\end{center}

\begin{center}
\begin{figure}[t]
\includegraphics[width=\columnwidth]{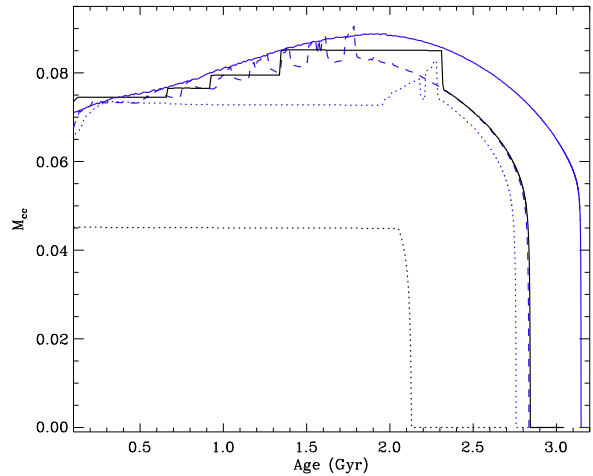}
\caption{Convective core evolution for models of 1.5 M$_\odot$. Black lines present the cases of Ledoux criterion with no semiconvective mixing (dots) and Schwarzschild criterion (solid). Blue lines show the models where additional mixing was included: semiconvection with $\alpha_{\mathrm{sc}}=0.001$ (dots), $\alpha_{\mathrm{sc}}=1$ (dashed), and overshooting (solid).} 
\label{fig:cores3}
\end{figure}
\end{center}

\begin{center}
\begin{figure}[t]
\includegraphics[width=\columnwidth]{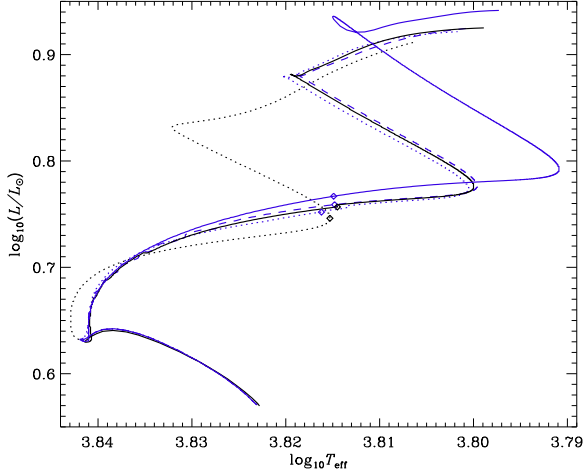}
\caption{Hertzsprung-Russell diagram for the 1.5 M$_\odot$ models. Black lines present the cases of Ledoux criterion with no semiconvective mixing (dots), and Schwarzschild criterion (solid). Blue lines show the models where additional mixing was included, semiconvection with $\alpha_{\mathrm{sc}}=0.001$ (dots), $\alpha_{\mathrm{sc}}=1$ (dashed), and overshooting (solid).} 
\label{fig:hrd}
\end{figure}
\end{center}

\begin{center}
\begin{figure}[t]
\includegraphics[width=\columnwidth]{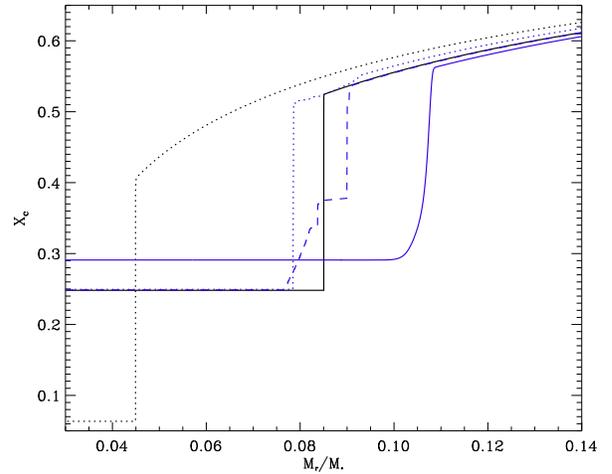}
\caption{Hydrogen profiles for the 1.5 M$_\odot$ models with Xc$\sim$0.25, marked with diamonds in Fig. \ref{fig:hrd}. Black lines present the cases of Ledoux criterion with no semiconvective mixing (dots), and Schwarzschild criterion (solid). Blue lines show the models where additional mixing was included, semiconvection with $\alpha_{\mathrm{sc}}=0.001$ (dots), $\alpha_{\mathrm{sc}}=1$ (dashed), and overshooting (solid).} 
\label{fig:chemical}
\end{figure}
\end{center}

For our model calculations we used the Garching Stellar Evolution Code (GARSTEC, \citet{ws08}). The input physics considered includes the OPAL equation of state \citep{fr96} complemented with the MHD equation of state for low temperatures \citep{hm88}, Ferguson's opacities for low temperature \citep{jf05}, OPAL opacities for high temperatures \citep{ir96}, the \citet{gs98} solar mixture, and the NACRE compilation for thermonuclear reaction rates \citep{ca99}. We computed models without diffusion and a He content of Y$=0.25$.\\
Several models were computed to test the different mixing prescriptions and the convective boundary definition. Within a convective zone the mixing is performed instantaneously, while in a semiconvective zone or overshooting region the mixing is carried out as explained in Sect. \ref{s:2}. Calculations were made for both the Schwarzschild and the Ledoux criterion for the definition of the convective zones, with and without including extra mixing due to overshooting and semiconvection. We explore the effects of this processes in models starting from the pre-main-sequence phase and evolved up to hydrogen exhaustion in the core.\\
In Fig. \ref{fig:cores}, we present the convective core evolution during the main-sequence lifetime for three models of 1.2, 1.5 and 2.0 M$_\odot$. The left panel shows the case where we applied the Schwarzschild criterion for convective boundaries determination, while the right panel shows the case where the Ledoux criterion was applied. In both cases, solid lines depict the size of the convective core, while the dotted lines represent the extent of the semiconvective region. No mixing was performed throughout the semiconvective zones ($\alpha_{\mathrm{sc}}=0$), which do appear in models for both convective criteria. We choose these models as they represent cases of interest: one model where the development of the convective core depends on the convective boundary definition used (1.2 M$_\odot$), one model with growing convective core during the main-sequence lifetime (1.5 M$_\odot$), and one model with a receding convective core during the H-burning phase (2.0 M$_\odot$). However, it is important to keep in mind that the effects of semiconvection are washed out when overshooting is included as described in Sect. \ref{s:2}, at least for the calibrated value of the efficiency parameter $\xi$, no matter which criterion is used to define the boundary of the convective regions. For the cases where it is present, the convective core disappears when the central hydrogen is exhausted, which corresponds to the end of the main-sequence phase. 
In this study, we will focus mainly on the 1.5 M$_\odot$ case, as it corresponds to the model which has the most extended semiconvective zone of all the considered ones. In Fig. \ref{fig:cores3} we present the main-sequence evolutionary models for this mass value, including now semiconvective mixing. As can be seen, already with a very small value for the efficiency parameter of semiconvection, this extra mixing makes the convective core grow much closer to the size of the Schwarzschild criterion case. This result is expected, as large values of $\alpha_{\mathrm{sc}}$ imply faster mixing, and in the limit when $\alpha_{\mathrm{sc}}\rightarrow\infty$ the mixing is performed instantaneously and the Schwarzschild limit should be recovered. As was already mentioned, we consider the main-sequence lifetime to end when hydrogen is exhausted in the center, which coincides with the center of the star becoming radiative. It is interesting to note that in our models, the lifetime on the main-sequence is reduced when the Ledoux criterion and semiconvective mixing are applied to the models with respect to the usual Schwarzschild criterion, which is an opposite result with respect to previous findings \citep[e.g.][]{cm82}. This is because in the previous implementations of semiconvection for low-mass stars, the mixing through that region recovering convective neutrality gave a fresh input of hydrogen fuel supply to the core and extended the H-burning phase. This is not the case when a diffusive, time-dependent mixing is considered for the semiconvective region as in our models.
In Fig. \ref{fig:hrd} we present a Hertzsprung-Russell diagram for the 1.5 M$_\odot$ models considered. Marked with diamonds are models which have very similar positions in this diagram, but the mixing prescriptions applied make their evolutionary stage in the main-sequence different. This is shown in Fig. \ref{fig:chemical}, where the hydrogen profiles are plotted for the selected models. The sharp chemical composition gradients are clearly seen, signature of the location of the convective core boundary. It is interesting to note that for the model computed with the Ledoux criterion and no mixing in the semiconvection region, the location in the HRD corresponds to a star almost at the end of its main-sequence lifetime, which is not the case for the other models considered being in a less advanced evolutionary stage.

\section{Conclusions and Future Prospects}\label{s:4}
In the present study, we have explored the effects of different convective boundary determinations and mixing prescriptions on the internal structure and evolution of low-mass stars. We have done so considering mainly the case of a 1.5 M$_\odot$ star, for which we have shown the difference in the internal structure resulting as a consequence of the applied physics, although the position in the Hertzsprung-Russel diagram remains very similar. In an upcoming study \citep{vs10}, we will further investigate the effects of these processes for different masses and metallicities, as well as changes in the efficiency parameters of semiconvective mixing and overshooting, and the inclusion of diffusive mixing. It is also important to keep in mind that semiconvection will play an important role in the evolution of a star depending on the transition between the \textit{pp} chain and the \textit{CNO} cycle as the main source of energy production (for a given metallicity). This transition is critically dependent on the value of the ($\rm{N}^{14} + p$) cross section, which has been subject of substantial reductions in the past years \citep[e.g.][]{mm08}.\\
Although accurate stellar parameters are of course of vital importance, asteroseismology can also be used as a tool to disentangle the observational degeneracy by focusing on the structural differences among the models produced by the mixing processes applied. It has been shown that both p-modes \citep{pd05} and g-modes \citep[e.g.][]{am08} are sensitive to changes of the molecular weight in the interior of stars, and could in principle be applied to distinguish the type of mixing that has occurred in the stellar interior.

\end{document}